\documentclass[11pt]{article}
\usepackage[latin1]{inputenc}
\usepackage[english]{babel}
\usepackage[namelimits]{amsmath}
\usepackage{amssymb}
\usepackage{amsmath}
\usepackage{amsthm}
\newcommand{\ac}{\'}

\begin{document}
\title{A LEMA\^{I}TRE-TOLMAN-FRIEDMANN UNIVERSE WITHOUT DARK ENERGY}
\author{Stefano Viaggiu
\\
Dipartimento di Matematica, Universit\'a "Tor Vergata"',\\ 
Via della Ricerca Scientifica, 1\\
Rome, Italy 00133,\\
viaggiu@axp.mat.uniroma2.it}
\date{\today}\maketitle
\begin{abstract}
We build a simple cosmological model by means of a parabolic 
Lema\^{i}tre-Tolman-Bondi (LTB) metric up to a redshift $z\simeq 0.4$,
an hyperbolic Friedmann metric for $z\sim 0.4$ up to the scale
where dimming galaxies are observed ($z\sim 1.4$) 
and a bulk spatially flat metric up to the last scattering
surface. Following Wiltshire, by taking into account the different
rate of clocks for an observer at the centre of a parabolic LTB spacetime
with respect to a one in the hyperbolic Friedmann metric, an "apparent"
negative deceleration parameter is perceived by the observer at the centre
of LTB, provided that all the regularity conditions are imposed and 
the past null sections of the LTB  and the hyperbolic 
Friedmann metrics are identified. As a result, a first order 
Hubble law emerges at low redshifts. A parameter $K$ arises
driving the deceleration parameter perceived by the central LTB
observer. Finally, we obtain that a negative value for the
deceleration parameter is compatible with the observed energy-density
at our present epoch.
\end{abstract}
\section{Introduction}
Supernovae type Ia (SNIa) observations of the past decade seem to indicate 
an accelerating universe (\cite{1,2}). In the standard approach with 
the Friedmann-Lema\^{i}tre models (FRLW),
an accelerating universe invokes the 
presence of a large amount of the so called dark energy. In the 
FRLW picture, this dark energy is given by the cosmological
constant. The dark energy represents a puzzle and perhaps the biggest problem
in the modern cosmology. In fact, 
a direct detection of a cosmological constant is
still lacking. In the last decade, many attempts have been made 
(see \cite{3,20} and references therein) to obtain physically sensible models
predicting a negative value for the deceleration parameter. Some authors
(see for example \cite{13,15,16,18,19,20}) showed that inhomogeneities can generate
an accelerating universe by using LTB metrics (see \cite{m1,21,22}), but several conditions 
must be imposed (see \cite{3,18}) in order to build regular physically viable models.
In particular, in \cite{17} it is shown that LTB metrics can 
mimic the distance-redshift
relation of the FRLW models at least at the third order in a series 
expansion with respect
to the redshift near the centre where the observer
is located. More generally, in the LTB
solutions, apparent acceleration in the redshift-distance relation seen by a central
observer can be shown to coexist with a volume average deceleration on a spacelike
hypersurface (see \cite{222})
An accelerating universe
can also be builded by averaging inhomogeneities (see \cite{3,4,5,6,7,8,9,10,11,12,17})
by means of the techniques depicted in \cite{5,6,7}. For a review on 
inhomogeneous cosmological models see \cite{d1,d2}.
Particularly interesting is the idea
developed in Wiltshire's papers \cite{9,12}. In these papers the dimming of the distant 
galaxies is interpreted as a "mirage" effect. This effect is due to the different rate
of cloks located in averaged not expanding galaxies, where the metric is spatially flat,
with respect to clocks in voids where the spatial curvature is negative.
To a negative spatial curvature can be associated a positive quasilocal
energy. This gravitational energy is not local, according to 
the strong equivalence principle. This approach seems to be very promising.\\
In this paper we adopt the point of view present in \cite{9,12}. In particular,
we are interested in models that take into account the observed
inhomogeneity of the universe at least up to a redshift $z\simeq 0.1$ by means
of LTB metrics. We apply the reasonings of \cite{11}, 
that, as claimed by the author, represent the first level step in solving
the fitting problem (see \cite{loc}).
We use for our local near universe
a parabolic LTB metric, instead of a "dressed" (see \cite{7})
hyperbolic Friedmann metric. In particular, in the spirit of \cite{11},
we obtain a simple (crude!) model. Our starting point is the consideration 
that 
dimming galaxies are located at hight redshift ($z\sim 0.5-1.4$) \cite{25}.
Therefore, we build a model where, up to $z\simeq 0.4$, the universe is filled
with a parabolic LTB metric, for $z\geq 0.4$ with an hyperbolic Friedmann metric
and for $z\geq 1.4$ with a bulk spatially flat Friedmann metric, according to WMAP
satellite data \cite{26}. Similarly to \cite{11}, we neglect the coupling of the 
dynamics of the different scales depicted above. In this way, the 
SNIa dimming galaxies are in a region where the clocks are ticking slowly with respect
to the ones of an observer located at the centre of a parabolic LTB metric.
Within this crude model, firstly we obtain (see \cite{14}) a linear Hubble law
for small redshifts. Further, we show that a negative central deceleration 
parameter arises, provided that the necessary 
boundary and regularity conditions are imposed.
Furthermore, we show that a negative central deceleration parameter emerges only for
values of the density compatible with the ones actually observed (underdensity).
Following Wiltshire, since the SNIa observations are made along the past light 
cone, we identify the null sections ($\theta,\phi =const.$) of the parabolic and 
the hyperbolic "dimming" zone.\\
It should be noticed that cosmological models with only two metrics can be
exhaustively found in \cite{ad}.\\ 
In section 2 we present the metrics composing the model together with the initial
an regularity conditions. In section 3 we present our model.
In section 4 we obtain the linear Hubble law for low redshifts.
In section 5 regular solutions are discussed. 
In section 6 we obtain a distance-redshift relation together
with the expression for the central deceleration parameter.
Section 7 collects some final remarks and conclusions.
Finally, the appendix is devoted to the study of the
matching conditions.

\section{Initial and regularity conditions}
The starting point of our simple model is the consideration
that the dimming galaxies have been found at hight redshifts
($z\sim 0.5-1.4$). Consequently, we assume that the universe with
$z\simeq 0.5$ up to some units, is represented by an hyperbolic
Friedmann metric with negative spatial curvature.
In appropriate coordinates, the metric can be put in the form
\begin{eqnarray}
& &ds_{F}^{2}=-dt^2+{\overline{a}(t)}^2[d{\eta}_{F}^{2}+
{\sinh}^2{\eta}_{F}\;d{\Omega}^2],\label{1}\\
& &{\overline{H}}_i t=
\frac{{\overline{\Omega}}_i}{2{(1-{\overline{\Omega}}_i)}^{\frac{3}{2}}}
\left(\sinh\xi-\xi\right),\nonumber\\
& &{\overline{a}}(t)=\frac{{\overline{a}}_i{\overline{\Omega}}_i}
{2(1-{\overline{\Omega}}_i)}\left(\cosh\xi-1\right),\nonumber\\ 
& &{\overline{a}(t)}^2{\overline{H}}^2(1-{\overline{\Omega}})=1,\nonumber
\end{eqnarray}
where ${\overline{\Omega}}_i$ is an initial density parameter,
${\overline{H}}_i$ an initial Hubble constant and 
${\overline{a}}_i$ an initial expansion factor to be specified.
An observer in the portion of universe given by (\ref{1}) measures
the observables by means of the comoving time $t$. 
With respect to  this time, an observer in (\ref{1}) measures 
an Hubble flow with a time dependent Hubble constant
$\overline{H}$ given by
\begin{equation}
\overline{H}=\frac{1}{\overline{a}}\frac{d\overline{a}}{dt}=
\frac{2{\overline{H}}_i\sinh\xi}{{\overline{\Omega}}_i(\cosh\xi-1)}
{\left(1-{\overline{\Omega}}_i\right)}^{\frac{3}{2}}.\label{2}
\end{equation}
In what follows we adopt the simplifying assumption
that the dynamics (see \cite{11}) of the pieces composing
our model are independent. This is a "crude" assumption, 
but nevertheless it gives the possibility to study
the role of inhomogeneities and structures in the 
distance-redshift relation. 
According to WMAP data \cite{26}, over the "dimming" zone,
the universe can be modelled with a bulk Friedmann metric with
zero spatial curvature, i.e.
\begin{eqnarray}
& &ds_{B}^{2}=-dt_{B}^{2}+a_{B}^{2}(t_B)(d{\eta}_{B}^{2}+{\eta}_{B}^{2}\;
d{\Omega}^2),\nonumber\\
& &a_B(t_B)=a_{Bi}{\left(\frac{t_B}{t_i}\right)}^{\frac{2}{3}},
\label{3}
\end{eqnarray}
where in (\ref{3}) we have assumed a dust model. With (\ref{3}),
the Hubble flow $H_B$ is given by
\begin{equation}
H_B=\frac{1}{a_B}\frac{da_B}{dt_B}=\frac{2}{3t_B}.
\label{4}
\end{equation}
Finally, we assume to live at the centre of a parabolic
(vanishing spatial curvature) LTB spacetime up to a maximum of
$z\simeq 0.4$. The metric is
\begin{equation}
ds_{TB}^{2}=-{dT}^2+R_{,\eta}^2d{\eta}^2+R^2d{\Omega}^2,\label{5}
\end{equation}
where, as usual:
\begin{eqnarray}
& &4\pi\rho=\frac{M_{,\eta}(\eta)}{R_{,\eta}R^2},\label{6}\\
& &R(T,\eta)={\left(\frac{9GM(\eta)}{2}\right)}^{\frac{1}{3}}
{\left[T-Y_{0}(\eta)\right]}^{\frac{2}{3}}\label{7},
\end{eqnarray}
and subindices with comma denote partial derivative.
The arbitrary function $Y_{0}(\eta)$ is often called 
"bang function" and is usually interpreted as a big-bang
singularity surface. The arbitrary function
$M(\eta)$ represents the 
gravitational mass inside a volume of radius
$\eta$ and $\rho$ the local density.\\
Regularity conditions must be imposed to (\ref{5})-(\ref{7}).
First of all, the density (\ref{6}) must be positive everywhere
and finite at the centre $\eta=0$ where the observer is placed
and $R(T,\eta)$ must be vanishing at $\eta=0,\;\forall\;T$.\\ 
Mathematically (see \cite{14,C2}:
\begin{eqnarray}
& &\rho(T,\eta\rightarrow 0)=finite,\label{8}\\
& &R(T,\eta\rightarrow 0)\sim \eta f(t)\;\;,\;\;
R_{,\eta}>0\;\;\forall\;\eta ,t,\label{9}\\
& &R(\eta,T)>0\;\;\forall\;T\;\;and\;\;{\eta}>0\;\;,\;\;
M(\eta\rightarrow 0)\sim {\eta}^3.\label{10}
\end{eqnarray}
Further, no trapped shell singularities must arise
in the TB zone, i.e.
\begin{equation}
\eta >2G M(\eta).
\label{11}
\end{equation}
The absence of trapped shells is automatically satisfied near the centre,
provided that conditions (\ref{9})-(\ref{10}) are imposed.
We also impose (see \cite{18}) the condition
\begin{equation}
Y_{0,\eta}(\eta=0)=0,
\label{12}
\end{equation}
to avoid potential problems at the centre, such as the "weak
singularity" discussed in \cite{18}, although
the severity of this problem is debated 
(see \cite{2000, altr}).
In section 6 we show that the presence of this "weak singularity"
does not affect the central deceleration parameter.\\
We consider now the initial conditions that we must impose to
the metrics (\ref{1}), (\ref{3}) and (\ref{5}).
If we consider (see \cite{11}) an early time such that
${\overline{\Omega}}_i$ is close to unity, all the three 
scales depicted above must be matched at that early time.
As a result ${\overline{a}}_i\simeq a_{Bi}$.\\
The same condition for (\ref{7}) it gives
$R(T_i,\eta)=\eta {\overline{a}}_i$. This condition,
in terms of the bang function $Y_0$ reads:
\begin{equation}
Y_0(\eta)=T_i-\frac{{{\overline{a}}_i}^{\frac{3}{2}}{\eta}^{\frac{3}{2}}}
{3}\sqrt{\frac{2}{GM}}.
\label{13}
\end{equation}
To study the condition (\ref{12}) is more useful to write 
${Y}_0(\eta)$ as follows:
\begin{equation}
{Y}_0(\eta)=T-\frac{\sqrt{2}}{3}\frac{R^{\frac{3}{2}}}{\sqrt{GM}}.
\label{14}
\end{equation}
In this section, to represent the dimming "zone", we used the
hyperbolic Friedmann solution. This choice leads to simple
computations, while an hyperpolic LTB spacetime
generally does not allow to simple analytic expressions.
For example, the volume average (\ref{15}) 
for unbound LTB metrics is generally
not at our disposal in an explicit workable form (see \cite{3}). 
Further, note that the approximation used in this paper can be
justified both physically and mathematically. In the appendix
we show that the continuity of the first fundamental form
can be achivied on a non-comoving thin shell, ensuring that the
spacetime is connected. Further,
the Stephani metric (see \cite{ext}) could be used as a comoving
thick shell located between the two transition zones composing our model
(see the appendix).  
It is worth to be noticed
that to calculate the distance-redshift relation only the conformal null sections
of the metric come in action, since the astrophysical observations are
performed along the past null cone. 

\section{The model}
The further step in our study is to write the relations between the different scales 
composing the model. To this purpose, it is observed a broadly uniform 
Hubble law (see \cite{12} and references therein).
Therefore, following Wiltshire, we impose the equality of the Hubble flow of
the three scales. 
Concerning the metric (\ref{5}), we can calculate, in the spirit 
of the Buchert scheme \cite{5}, a volume average expansion up to 
some scale ${\eta}_D$, where dimming galaxies come in action.
For the proper volume $V_D$ and the expansion $\theta$, we read:
\begin{eqnarray}
& &V_D=4\pi{\int}_{0}^{{\eta}_D}R_{,\eta}R^2d\eta,\label{15}\\
& &{<\theta >}_D=\frac{d}{dT}\left(\ln V_D\right),\;
M_D=M({\eta}_D),\label{16}\\
& &a_D(T)=
{\left[\frac{(T-Y_0({\eta}_D))}{(T_i-Y_0({\eta}_D))}\right]}^{\frac{2}{3}}.
\label{18}
\end{eqnarray}
As a result, for the averaged Hubble flow 
$T_{TB}$ in the LTB sector, we have 
\begin{equation}
H_{TB}=\frac{{<\theta >}_D}{3}=\frac{2}{3[T-Y_0({\eta}_D)]}.
\label{19}
\end{equation}
After equating (\ref{2}), (\ref{4}) and (\ref{19}), we get:
\begin{eqnarray}
& &t_B=T-Y_0({\eta}_D),\label{20}\\
& &T({\xi})=\frac{{\overline{\Omega}}_i}{3
{{\overline{H}}_i(1-{\overline{\Omega}}_i)}^{\frac{3}{2}}}
\frac{{\left(\cosh\xi-1\right)}^2}{\sinh\xi}+Y_0({\eta}_D).\label{21}
\end{eqnarray}
Apart from the constant in the right hand side, the expression (\ref{21})
for the time delay is the one found in \cite{11}. This is not a surprise 
because, after averaging a parabolic LTB metric, we obtain an
Hubble flow, equation (\ref{19}), that, apart from the constant
$Y_0({\eta}_D)$, is the same of the bulk spatially flat metric.
The constant $Y_0({\eta}_D)$ takes into account the cut-off made in
our model. Further, the term $Y_0({\eta}_D)$ appears as a translational
factor and thus does not enter in our analysis. We only mention the 
fact that such a constant can give a correction (positive if
$Y_0({\eta}_D)>0$) to the age of the universe. For the 
lapse function $J(\xi)=\frac{dt}{dT}$, we get:
\begin{equation}
J(\xi)=\frac{3}{2}\frac{(1+\cosh\xi)}{(2+\cosh\xi)}.
\label{22}
\end{equation}
Formulas (\ref{21})-(\ref{22}) describe the different rate of clocks between 
the parabolic TB observer and an hypothetical observer
placed where the galaxies are dimming. 
Obviously, for the reasonings above, formula (\ref{22}) is 
exactly the one found in \cite{11}, but
is expressed in a different background.\\ 
If we want to describe the dimming of distant
galaxies, we must to relate the metrics
(\ref{1}) and (\ref{5}) on the past null cone, where
the SNIa observations are performed. To this purpose,
the radial null sections ($\theta,\phi=const.$) of (\ref{1})
and (\ref{5}) must be the same, i.e. 
$ds_{F}^2=J^2ds_{TB}^2$. As a result, the following equations
hold on inward radial null geodesics:
\begin{eqnarray}
& &\frac{J}{\overline{a}}\;dT=-d{\eta}_F,\label{23}\\
& &R_{,\eta}\;d\eta = -dT.\label{24}
\end{eqnarray}
Therefore, our related metric in the LTB inhomogeneous spacetime is
\begin{equation}
ds_{TB}^{2}=-{dT}^2+\frac{{\overline{a}}^2}{J^2}d{\eta}_{F}^{2}+
R^2d{\Omega}^2.
\label{25}
\end{equation}
For a more complete discussion regarding the matching conditions
see the appendix.\\
For the metric (\ref{25}), we can define a radial observed
(by the central observer) Hubble flow with
\begin{equation}
H_{ob}=\frac{1}{a}\frac{da}{dT}=\frac{d}{dT}\left(\ln\frac{\overline{a}}{J}\right).
\label{26}
\end{equation}
It is in terms of (\ref{26}) that we measure the Hubble flow. Concerning the 
luminosity-distance $d_L(z)$, for the metric (\ref{25}) we get (see 
\cite{16,23,24,dl1,dl2,dl3}):
\begin{equation}
d_L=B_0{(1+z)}^2\;,\;B_0^2=\frac{dS_0}{d{\Omega}_0}=R^2,
\label{27}
\end{equation}
where ${\Omega}_0$ is the solid angle subtended by a bundle of
null geodesics
diverging from the observer and $S_0$ is the cross-sectional
area of the bundle.
We must to integrate the relevant field equations for our 
purpose. First of all, by means of (\ref{1}), (\ref{22})
we can integrate equation (\ref{23}) along the past null cone.
We get:
\begin{equation}
{\eta}_F={\xi}_0-\xi\;\;,\;\;\xi\leq {\xi}_0,
\label{27b}
\end{equation}
where the subscript "$0$" denotes the actual time  related to the central
TB observer. Following C{\ac{e}}l{\ac{e}}rier \cite{16}, for the metric
(\ref{25}) we can express the observed redshift $z$ in terms of the time
parameter $\xi$, i.e.:
\begin{equation}
\frac{d{\eta}_F}{dz}=\frac{1}{(1+z)}\frac{1}
{\frac{d}{dT}\left(\frac{\overline{a}}{J}\right)},
\label{28}
\end{equation}
and therefore, by means of (\ref{23}) and integrating backward starting 
from $z=0$, we read:
\begin{equation}
1+z=\frac{J}{J_0}\frac{{\overline{a}}_0}{\overline{a}}.
\label{29}
\end{equation}
Since the expression (\ref{29}) has been evaluated along
the past null cone, it is the same found in \cite{11}.
Nevertheless, since inhomogeneities have been taken into
account in our model, 
we expect corrections with respect to the picture of 
the paper \cite{11}, in particular in the relation distance-redshift
$d_L(z)$. In evaluating the functions entering in the relation
$d_L(z)$, we need of the difference $T_0-T$, with $T_0$ the 
actual time. The equation (\ref{29}) permit us to express 
$\cosh\xi$ in terms of the redshift $z$. As a result, after 
noticing that ${\overline{\Omega}}_0=\frac{2}{1+\cosh{\xi}_0}$
and with the help of (\ref{1}) and (\ref{21}), we obtain:
\begin{eqnarray}
& &T_0-T=
A\left[\frac{{(\cosh{\xi}_0-1)}^{\frac{3}{2}}}{{(1+\cosh{\xi}_0)}^{\frac{1}{2}}}-
\frac{{(\cosh{\xi}-1)}^{\frac{3}{2}}}{{(1+\cosh{\xi})}^{\frac{1}{2}}}\right],\nonumber\\
& &A=\frac{{\overline{\Omega}}_0(2+{{\overline{\Omega}}_0}^2)}
{H_0{(1-{\overline{\Omega}}_0)}^{\frac{3}{2}}}\frac{1}{{(2+{\overline{\Omega}}_0)}^2},
\label{30}
\end{eqnarray}
where $H_0$ is the measured Hubble constant given by (\ref{26}) and calculated
at the present time ${\xi}_0$ (or $T_0$), i.e.
\begin{equation}
H_0=3{\overline{H}}_0
\left[\frac{2+{{\overline{\Omega}}_0}^2}{{(2+{\overline{\Omega}}_0)}^2}\right].
\label{ale}
\end{equation}
Finally, along the past null geodesics 
we have $M_{,\eta}=M_{,T}\frac{dT}{d\eta}$, and  thus, thanks to (\ref{24}), expression
(\ref{6}) becomes:
\begin{equation}
{\rho}(T)=-\frac{M_{,T}}{4\pi R(T)^2}.
\label{30bis}
\end{equation}
Obviously, regularity of (\ref{30bis}) for $T\rightarrow T_0$ requires that:
\begin{eqnarray}
& &M(T\rightarrow T_0)\;\sim\;{(T_0-T)}^3+o(1),\label{31}\\
& &R(T\rightarrow T_0)\;\sim\;(T_0-T)+o(1).\nonumber
\end{eqnarray}

\section{Zeroth order solution: the linear Hubble law}
To complete our model, we must integrate the equation (\ref{24}).
Obviously, because of
the partial derivative of $R$ in the left hand side,
this equation cannot be integrated in this form.
Nevertheless, the equation (\ref{24}) can be easily integrated as follows.
Firstly, we write:
\begin{equation}
\left(\frac{(dR(\eta,T(\eta))}{d\eta}\right)d\eta=
\left(\frac{dR(T,\eta(T))}{dT}\right)dT=dR=
R_{,\eta}\;d\eta+R_{,T}\;dT,
\label{32}
\end{equation}
and the equation (\ref{24}) becomes
\begin{equation}
dT=\frac{dR}{(R_{,T}-1)}.
\label{32}
\end{equation}
Further, from (\ref{7}) we obtain
$R_{,T}=\sqrt{\frac{2GM}{R}}$, and as a result
\begin{equation}
dT=\frac{dR}{\left(\sqrt{\frac{2GM}{R}}-1\right)}.
\label{34}
\end{equation}
We will explain the general strategy to integrate the equation (\ref{34})
in the next section. First of all we are interested in 
the first order calculation of $d_L(z)$. The first condition to impose 
is $R>2GM$, that is equivalent, near the centre, to the first of
conditions (\ref{11}) and ruls out trapped shell singularities.
It is worth to noticing that the conditions (\ref{31}) near the centre are sufficient
to satisfy the condition above mentioned.
The zeroth order approximation for (\ref{34}) emerges when $R>>2GM$. This extreme 
approximation it gives the correct first order of $d_L(z)$. Consequently,
after integrating with the appropriate boundary condition
($R(T_0)=0$), we have:
\begin{equation}
R(T,\eta(T))=R(T)=T_0-T+o(1).
\label{35}
\end{equation}
With the help of (\ref{29}), (\ref{30}), we read:
\begin{equation}
d_L(z)=\frac{z}{H_0}+o(z),
\label{36}
\end{equation}
that is the well known linear Hubble law for low redshifts.
Therefore, the zeroth order of our model it gives the observed
distance-redshift relation for $z<<1$.

\section{Exact regular solutions}
We can rewrite equation (\ref{34}) as:
\begin{equation}
R(T)=-\int_{T_0}^{T}\left(1-\sqrt{\frac{2GM}{R}}\right)dT.
\label{37}
\end{equation}
To integrate (\ref{37}), we write $\sqrt{\frac{2GM}{R}}=F(T)$, 
being $F(T)$ a regular differentiable function. With the conditions
(\ref{31}) we must impose:
\begin{equation}
F(T)\in (0,1)\;\;,\;\;R(T)>0.
\label{HH}
\end{equation}
In this way, after fixing an ansatz for $F(
T)$, we can integrate the equation (\ref{37}),
obtaining:
\begin{equation}
R(T)=T_0-T+\int_{T_0}^{T} F(T)dT.
\label{38}
\end{equation}
After solving the equation (\ref{38}) for $R(T)$, $M(T)$ is given by
\begin{equation}
M(T)=\frac{R}{2G}F(T)^2.
\label{39}
\end{equation}
Further, we impose the condition (\ref{12}), that in terms of equation
(\ref{14}) seen as a function of $T$ becomes:
\begin{equation}
Y_{0,T}(T_0)=0.
\label{40}
\end{equation}
Finally, the relation between $\eta$ and $T$ along the past null cone
is obtained by inverting the equation (\ref{7}) with the 
help of (\ref{13}) i.e.
\begin{equation}
\eta={\left(\frac{9GM}{2{{\overline{a}}_i}^3}\right)}^{\frac{1}{3}}
{\left[R^{\frac{3}{2}}\sqrt{\frac{2}{9GM}}-T+T_i\right]}^{\frac{2}{3}}.
\label{41}
\end{equation}
In the following (in particular for the distance-redshift
relation (\ref{44})) it is essential the behaviour of $F(T)$
(and $R(T)$) near the observer at the centre. As a result,
any given expression for $F(T)$ must have a taylor expansion near the 
centre fixed by the regularity conditions (\ref{HH}) and
(\ref{40}). Hence, we can take for 
$F(T)$ a polynomial expression.
Therefore, for our purposes,
without loss of generality, we can take  for $T\leq T_0$
\begin{eqnarray}
& &F=\frac{H_0}{K}(T_0-T)+\frac{{H}_{0}^{2}}{K^2}{(T_0-T)}^2+
Q{H}_{0}^{3}{(T_0-T)}^3\label{42}\\
& &R=T_0-T-\frac{H_0}{2K}{(T_0-T)}^2-\frac{{H}_{0}^{2}}{3K^2}{(T_0-T)}^3-\nonumber\\
& &-\frac{Q}{4}{H}_{0}^{3}{(T_0-T)}^4,\label{43}
\end{eqnarray}
with $K$ and $Q$ adimensional constant. It is worth to be noticed that 
the first two terms in (\ref{42}) are fixed by the condition (\ref{40}).
If we impose that the LTB scale is extended up to a maximum of 
$z\simeq 0.4$, we
see that the conditions (\ref{HH}) are satisfied for $K>\frac{1}{2}$, with 
$Q$ of the same order of $K$. Further, note that the limit
$z\simeq 0.4$ for the boundary of the TB metric only changes the 
allowed values for $K$ and $Q$. For example, for $z<0.4$
$K>a>\frac{1}{2}$.
Furthermore, note that if we impose the condition $Y_0\geq 0$
for $T\rightarrow T_0$ with $Y_0(T_0)=0$, we must have:
\begin{equation}
K=\frac{3}{2}H_0 T_0,
\label{af}
\end{equation}
while, if we take $K<\frac{3}{2}H_0 T_0$ (but positive), then $Y_0\geq 0$
with $Y_0(T_0)\neq 0$.
As we see in the next section, the positivity of $Y_0$, i.e. condition
(\ref{af}) implies a maximum possible value for the central deceleration
parameter.
It should be noticed that if we take a more general 
expression other than (\ref{42}), in order to satisfy all the regularity
conditions depicted above,
the Taylor expansion
of $F(T)$ near the centre 
must be equal to expression (\ref{42}), at least for the first two terms.
In the next section we show that are exactly these terms that enter in the 
expression for the central deceleration parameter. 
As a final remark for this section, note that
in our model, thanks to the equation
(\ref{39}), the function $F(T)$ is related to $M(T)$  on the light null cone. 
Further, by means of the equation
(\ref{41}) we can (at least in principle) find $M=M(\eta)$, i.e.
the dependence in terms of $\eta$. Consequently, 
with respect to our construction,
the FLRW limit can be obtained by setting the particular expression for 
$F(T)$ on the past null cone such that, when expressed in terms of
$\eta$ by means of the equation (\ref{41}), we have
$M(\eta)\sim{\eta}^3$ or $Y_0(\eta)=constant$. 

\section{Central deceleration parameter and observed density} 
With the help of (\ref{29}), (\ref{30}) we can express $M(T), R(T)$
in terms of the measured redshift $z$. By taking the expressions
(\ref{42}), (\ref{43}) and after a Taylor expansion near the centre
($z=0$), we get:
\begin{eqnarray}
& &d_L=\frac{z}{H_0}+\label{44}\\
& &+\frac{z^2}{4H_0 K{(2+{\overline{\Omega}}_{0}^{2})}^2}
\left[5K{\overline{\Omega}}_{0}^{4}-2{\overline{\Omega}}_{0}^{4}+4K{\overline{\Omega}}_{0}^{3}
-8{\overline{\Omega}}_{0}^{2}+2K{\overline{\Omega}}_{0}^{2}-8+16K\right]+o(z^2).
\nonumber
\end{eqnarray} 
The central deceleration parameter $q_0$ is given by (see \cite{16,18,27})
\begin{equation}
q_0=-H_0\frac{d^2}{dz^2}\left(d_L(z=0)\right)+1,
\label{45}
\end{equation}
that with (\ref{44}) becomes:
\begin{equation}
q_0=\frac{\left[-3K{\overline{\Omega}}_{0}^{4}+2{\overline{\Omega}}_{0}^{4}-
4K{\overline{\Omega}}_{0}^{3}+8{\overline{\Omega}}_{0}^{2}+6K{\overline{\Omega}}_{0}^{2}+
8-8K\right]}{2K{(2+{\overline{\Omega}}_{0}^{2})}^2}.
\label{46}
\end{equation}
If we do not consider the flat Friedmann metric, we
could to estimate
the bulk central deceleration parameter $q_{0B}$
at early times 
by setting ${\overline{\Omega}}_{0}=1$ in (\ref{46}). As a result:
\begin{equation}
q_{0B}=-\frac{1}{2}+\frac{1}{K}.
\label{47}
\end{equation}
Note that for $K\rightarrow 1$ it follows that $q_{0B}\simeq\frac{1}{2}$.\\
By taking the asymptotic limit $T_0\rightarrow\infty$
(${\overline{\Omega}}_{0}\rightarrow 0$) we have 
\begin{equation}
q_{0\infty}=-1+\frac{1}{K}.
\label{48}
\end{equation}
From (\ref{48}) we see that our model is consistent
with  an accelerating universe
provided that $K>1$. For very large values of $K$ we obtain
$q_{0\infty}\simeq -1$.
If we put in (\ref{46}) an estimate value for
${\overline{\Omega}}_{0}$ as, for example, ${\overline{\Omega}}_{0}\leq\frac{1}{10}$,
we obtain that for $K\rightarrow 1^{+}$ again $q_0<0$. More generally, 
$q_0 < 0$ at the times for which:
\begin{equation}
K>\frac{(2{\overline{\Omega}}_{0}^{4}+8{\overline{\Omega}}_{0}^{2}+8)}
{(3{\overline{\Omega}}_{0}^{4}+4{\overline{\Omega}}_{0}^{3}-6{\overline{\Omega}}_{0}^{2}+8)}.
\label{49}
\end{equation}
For example, for ${\overline{\Omega}}_{0}\simeq\frac{1}{10}$, 
$K>1.0017$. Therefore, also for the actual universe our model admits
a negative central deceleration parameter. By taking the limit value for $K$ given by
(\ref{af}), ($Y_{0}(T_0)=0$) we obtain a maximum negative value for 
$q_{0\infty}$ i.e. $q_{0\infty}=-\frac{1}{3}$. 
As a final step, we consider the local density given by (\ref{30bis}).
It is a simple matter to see that, with expressions (\ref{42}), (\ref{43}),
by taking the limit $T\rightarrow T_0$, we obtain:
\begin{equation}
{\rho}_0=\rho (T_0)=\frac{3H_{0}^{2}}{8\pi GK^2}.
\label{50}
\end{equation}
Expression (\ref{50}) represents an interesting result:
an actual underdensity is in agreement with a negative
value for $q_0$. This is in agreement with the actual exstimation
predicted by Friedmann models with ${\rho}_c=\frac{3H_{0}^{2}}{8\pi G}$
and $\frac{\rho}{{\rho}_c}<1$.\\
Finally, note that our calculations can be easily changed if we
do not impose the condition (\ref{12}). We must only make the following 
substitutions in the equation (\ref{42}): $K\rightarrow a$,
$K^2\rightarrow b$, being $(a,b)>0$ (the weak singularity is ruled out
when $a^2=b$). As a result, since the $q_0$ parameter only involves
the second order in $T_0-T$ in the expression (\ref{43}),
the central deceleration parameter remains unaffected and so 
also expression (\ref{50}) does not change. The so called weak 
singularity comes in action only at the third order in $z$ in the
redshift-distance relation.

\section{Conclusions}
Following Wiltshire papers, we build a model for the universe without 
dark energy, by taking into account the observed inhomogeneous universe
for low redshifts by means of a LTB metric. When the clocks of the 
observer at the centre of a parabolic LTB spacetime are
related to the ones placed in an hyperbolic Friedmann metric where
are located the dimming galaxies, an apparent negative value for
the central deceleration parameter arises, provided that
all the regularity and boundary conditions are imposed.
We attempted to introduce a model in which both a description
of the low redshift irregularities and of the large scale homogeneity
where present. Since a coupling between the dynamics of the different scales
composing the universe is neglected, the model is a "crude"
approximation. Nevertheless, our model allows for a negative value of the
central deceleration parameter that is in agreement with the observed
underdensity for the actual universe \cite{26}. With respect to
the Friedmannian picture of \cite{11}, the use of a LTB parabolic metric 
to describe our nearby universe, permit us the introduction of a
dimensionless parameter $K$ whose allowed values are in agreement
with an accelerating universe. In \cite{11} the deceleration parameter
runs to zero asymptotically from positive values. \\
In any case, we have also shown that the so called weak singularity
comes in action only at the third order in $z$ in the expression
(\ref{44}), and as a result the parameter $q_0$ does not
"feel" such a hypothetical singularity.\\
Physically, at the scale where dimming galaxies come in action,
we adopt an hyperbolic Friedmann metric. In fact, 
with such a metric is associated a positive gravitational 
energy that is the "source" of the slowing rate of the clocks when
compareted with the central (parabolic) LTB observer. 
According to WMAP satellite data \cite{26} and CMBR \cite{28}
over the "dimming" zone, we adopt a spatially flat Friedmann
metric. As a final remark we cite the recent paper
\cite{pro} where, by means of an 
investigation of new sperimental data,
it seems that we live in a universe with a deceleration 
parameter near to $q_0\simeq 0^{-}$, i.e. with a slowing down
of the cosmic acceleration. These results seem to be in disagreement
with the standard LCDM model and could encourage
the point of view of our work.\\
A further development of this paper could be 
to consider, as in \cite{10,12}, the dynamics by means of 
the full Buchert \cite{5} formalism by taking into account 
the inhomogeneous structure of the nearby observed universe
together with the backreaction or to use the full covariant
machine given in \cite{6}.
Not a simple task!
\section*{Acknowledgments}
I would like to thank Luciano Pietronero for hints and suggestions.\\
I would also like to thank Andrea Lionetto,
Francesco Sylos Labini and Giuseppe Ruzzi for
useful discussions. 
\section*{APPENDIX}
We now study the matching problem between a parabolic LTB solution and
an hyperbolic Friedmann solution on a regular comoving surface 
$S$. Since of the spherical symmetry of both metrics, we can use
spherical symmetric coordinates ${\xi}^{\alpha}$ on $S$. Therefore,
if we denote with ${\psi}^{\alpha}_{+}$ the coordinates of the Friedmann
metric 
and with ${\psi}^{\alpha}_{-}$ the ones of the LTB metric, we have
(see \cite{23}) ${\xi}^{\alpha}={\psi}^{\alpha}_{+}={\psi}^{\alpha}_{-}$
and, as a result, we can take ${\xi}^{0}=t=T,\;{\xi}^{2}=\theta,\;
{\xi}^{3}=\phi$. In this way the comoving surface $S$ is given by:
\begin{equation}
S_{-}=\eta-S_{0}=0\;\;,\;\;S_{+}={\eta}_F-S_{0}=0,
\label{A1}
\end{equation}
where $S_{0}$ denotes the boundary of the LTB metric.
The continuity of the first fundamental form (see \cite{f1,f2})
on $S$, i.e. $ds^{2}_{S}=g_{\alpha\beta}d{\xi}^{\alpha}d{\xi}^{\beta}$
($\alpha =(0,2,3)$) leads to $ds^{2}_{-}=ds^{2}_{+}$, i.e.
\begin{equation}
{(R)}_{|S}={\overline{a}}{(\sinh{\eta}_F)}_{|S}.
\label{A2}
\end{equation}
For the unit normals $n^{-}_{\mu}$ and $n^{+}_{\mu}$ we have
\begin{equation}
n^{-}_{\mu}=R_{,\eta}{\delta}^{1}_{\mu}\;\;,\;\;
n^{+}_{\mu}={\overline{a}}{\delta}^{1}_{\mu}.
\label{A3}
\end{equation}
For the second fundamental form 
$K_{\alpha\beta}={(n_{\alpha ;\beta})}_{|S}$ we have
$K_{\alpha\beta}=\frac{\partial{\psi}_{\mu}}{\partial{\xi}^{\alpha}}
\frac{\partial{\psi}_{\nu}}{\partial{\xi}^{\beta}}K_{\mu\nu}$.
The continuity condition on $S$, i.e. 
$K^{-}_{\alpha\beta}=K^{+}_{\alpha\beta}$, becomes
\begin{equation}
1={(\cosh{\eta}_F)}_{|S}.
\label{A4}
\end{equation}
From equations (\ref{A2}) and (\ref{A4}) it is evident that is not formally possible
to match an hyperbolic Friedmann metric with a parabolic LTB one on the
comoving surface (\ref{A1}).\\ 
If we take for the outer metric ($+$) an unbound
comoving LTB line element
\begin{equation}
ds^{2}_{TB}=-d{\tilde{T}}^2+\frac{{\tilde{R}}^{2}_{,\eta}}{f^2}d{\eta}^2+
{\tilde{R}}^2 d{\Omega}^2,
\label{A5}
\end{equation}
with $f^2>1$, insted of (\ref{A4}) we have
\begin{equation}
{(R)}_{|S}={(\tilde{R})}_{|S}\;,\;T={\tilde{T}}+\;\;constant\;,\;
{(f)}_{|S}=1.
\label{A6}
\end{equation}
As a result, for the lapse function $J$ we have
$J=\frac{dT}{d{\tilde{T}}}=1$.\\
Remember that, since astrophysical observations seem to 
show a broadly uniform "Hubble flow"
(see \cite{12}), we must perform the
matching according to this astrophysical evidence.
As a result, the equation (\ref{22})
must be satisfied for the lapse function ($J\neq 1$).
Hence, as an example,
we can think to match on a non-comoving thin shell.
To do this, we can take for the shell:
\begin{equation}
\eta={\eta}_s(T(\tau)),\;\;\;{\eta}_F={\eta}_{fs}(t(\tau)),
\label{A10}
\end{equation}
where $\tau$ is the proper time on the shell.
Since we have $T=T(\tau),\;t=t(\tau)$, the continuity
of the pull-back of the metric on the non-comoving shell
$S$ it gives (on $S$):
\begin{eqnarray}
& &\frac{dT}{d\tau}=A,\label{A101}\\
& &\frac{dt}{d\tau}=B,\label{A102}\\
& &\overline{a}\sinh {\eta}_{fs}=R,\label{A103}\\
& &\frac{dt}{dT}=J,\label{A104}\\
& &A=\sqrt{1+R_{,\eta}^{2}{\dot{\eta}}_{s}^{2}},\label{A1004}\\
& &B=\sqrt{1+{\overline{a}}^2{\dot{\eta}}_{fs}^{2}},\label{A10004}
\end{eqnarray}
where dot is the derivative of the proper time on the shell
(all the expressions are calculated on the surface (\ref{A10})).
The system (\ref{A101})-(\ref{A10004}) can be integrated in
different ways. As an example, from the equation (\ref{A101}) we have
\begin{equation}
d\tau=\sqrt{1-{\left(\frac{d{\eta}_s}{dT}\right)}^2
R_{,\eta}^{2}}dT.
\label{A105}
\end{equation}
Thanks to (\ref{A105}), equation (\ref{A102}) becomes:
\begin{equation}
J^2-1=-{\left(\frac{d{\eta}_s}{dT}\right)}^2
R_{,\eta}^{2}+{\overline{a}}^2{\left(\frac{d{\eta}_{fs}}{dT}\right)}^2.
\label{A106}
\end{equation}
The equation (\ref{A103}) permit us to express
$\sinh{\eta}_{fs}$ in terms of $({\eta}_{s}, T)$, and thus
we can use equation (\ref{A106}) to resolve for ${\eta}_{s}$.
Finally, we can integrate equation (\ref{A105}) to obtain the
relation $\tau=H(T)$.\\
For the unit normals we have:
\begin{eqnarray}
& &{\eta}_{\mu}^{-}=\left[-\dot{{\eta}_{s}}R_{,\eta},\;
R_{,\eta}A,\; 0,\; 0\right],\label{A107}\\
& &{\eta}_{\mu}^{+}=\left[-{\dot{\eta}}_{fs}\overline{a},\;
\overline{a} B,\; 0,\; 0\right].\label{A108}
\end{eqnarray}
The equations for the continuity of the extrinsic curvature are:
\begin{eqnarray}
& &\dot{{\eta}_{s}}R_{,\eta}RR_{,T}+RA={\dot{\eta}}_{fs}
{\overline{a}}^2{\overline{a}}_{,t}{\sinh}^2{\eta}_{fs},+\nonumber\\
& &\overline{a} B\sinh{{\eta}_{fs}}\cosh{{\eta}_{fs}},\label{A109}\\
& &\dot{{\eta}_{s}}R_{,\eta}{\ddot{T}}-R_{,\eta}{\ddot{{\eta}_{s}}}A+\nonumber\\
& &{\dot{\eta}}_{s}^{3}R_{,\eta}^{2}R_{,\eta,T}-
R_{,\eta,\eta}{\dot{\eta}}_{s}^{2}A=\nonumber\\
& &{\dot{\eta}}_{fs}\overline{a}\ddot{t}-\overline{a}B{\ddot{\eta}}_{fs}+
{\dot{\eta}}_{fs}^{3}{\overline{a}}^2{\overline{a}}_{,t}.\nonumber
\end{eqnarray}
Conditions (\ref{A109}) seem to be incompatible with the
equations (\ref{A101})-(\ref{A104}). In particular,
we have no sufficient number of functions to satisfy
the conditions (\ref{A109}).\\
Concerning the matching between the hyperbolic and the parobolic
Friedmann solutions, for the first fundamental form we have 
the equations (\ref{A101})-(\ref{A10004}) with 
$R_{,\eta}\rightarrow a_B$ in (\ref{A1004}), $R\rightarrow a_B$ 
in (\ref{A103}) and
$J=\frac{dt}{dt_B}$ instead of the equation (\ref{A104})
(with the same $J$!)
and, obviously, with different
non-comoving surfaces and proper time on the shell. 
Hence, also for the Friedmann metrics 
all the reasonings after equation (\ref{A10004}) are also valid.
As a result, at least with respect to the
intrinsic curvature, the matching conditions can be satisfied. 
It is worth to be noticed that the continuity of the metric on 
the non-comoving thin shell is the minimum requirement ensuring that
the whole spacetime is connected.\\
We briefly discuss another possibility. We can take the 
spherically simmetric perfect fluid Stephani metric (see \cite{ext}):
\begin{eqnarray}
& &ds^2=-D^2d{\tau}^2+\frac{Y^2}{V^2}\left[dr^2+r^2 d{\Omega}^2\right],
\nonumber\\
& &V=1+\frac{1}{4}k(\tau)r^2,\nonumber\\
& &D(\tau,r)=F(\tau)\frac{Y(\tau)}{V}\frac{d}{d\tau}\left(\frac{V}{Y}\right),
\nonumber\\
& &k(\tau)=Y^2\left[C^2(\tau)-\frac{1}{F^2(\tau)}\right],\label{Aba}
\end{eqnarray}
being $F,C,Y$ free functions.
The metric (\ref{Aba}) can be used to model a comoving
thick shell (see \cite{ado})
between the LTB and the hyperbolic Friedmann zone and so also between the
hyperbolic and the parabolic Friedmann metrics.
Differently from the thin shell discussed above,
the metric (\ref{Aba}) has, at least in principle,
a sufficient number of arbitrary
functions to perform the matching by imposing the continuity
of the first and the second fundamental form, avoiding 
surface-layer matter, althought in practice this can result not easy.
However, in this case the calculations of this article could 
be changed. In any case, 
if this thick shell is "small" with respect to the glued
regions, one may believe that the deviation from the calculations
performed in this paper remains "small". In particular, if we denote with
${\eta}_1, {\eta}_2$ the boundary of the thick shell between the parabolic
and the hyperbolic metric, then if ${\eta}_2-{\eta}_1$ 
is ``small'', we expect
a ``small'' deviation on the redshift-distance relation.
Obviously, we could use more general metrics for the thick shell
than the Stephani one. In a future work we shall consider the model
of this paper with the introduction of a comoving thick shell.\\
As a further remark for this appendix, 
it should be stressed that, from astrophysical data,
emerges the necessity of more scales to describe the whole
universe we observe. If this is the case, the observed broadly uniform 
Hubble flow imposes the equality of the (spatial averaged) Hubble flow
of the different metrics composing the universe we observe. As shown in this
paper (see equation (\ref{21}), the equality of the Hubble flow leads to 
a clock delay effect. If this reasoning is correct, as a consequence, the time flow
cannot be chosen globally uniform.\\
In this paper, the matching is performed by equating the conformally related
null sections with conformal factor given by $J$ (the lapse function).
It is worth to be noticed that for the calculation of the distance-redshift relation,
only the null geodesics come in action.  
Generally, since only null geodesics can probe the
cosmological scales under consideration, it seems to be reasonable to impose,
at least, the matching between the null sections. In this way we have a crude 
but sufficient approximation
to explore the role of the observed inhomogeneities
on the distance-redshift relation. As a final remark, note that, with the 
introduction of a LTB parabolic metric, we have shown that the calculation 
performed in \cite{11} are compatible with an accelerating universe, while 
in \cite{11} the use of only Friedmann spacetimes leads to a universe 
with $q\rightarrow 0^{+}$ at late times. 

\end{document}